\documentclass[preprint,12pt,3p]{elsarticle}

\usepackage{amssymb}
\usepackage{multirow}
\usepackage{color}
\usepackage{framed} 
\usepackage{multicol} 
\usepackage{graphicx}
\usepackage{subfigmat}
\usepackage{subfigure}
\usepackage{float}
\usepackage{amsmath,bm}
\usepackage[version=4]{mhchem}
\usepackage{siunitx}
\usepackage{natbib,hyperref}
\usepackage{longtable,tabularx}
\usepackage{setspace}

\usepackage{nomencl} 

\makeatletter
\def\blfootnote{\xdef\@thefnmark{}\@footnotetext}
\makenomenclature
\journal{Elsevier}

\begin{document}

\begin{frontmatter}

\title{Vehicle Platooning Impact on Drag Coefficients and Energy/Fuel Saving Implications}

\author{Ahmed A. Hussein \footnote{Post-Doctorate Fellow, Center for Sustainable Mobility, Virginia Tech Transportation Institute, 3500 Transportation Research Plaza (0536) Blacksburg, VA 24061, USA.}}
\author{Hesham A. Rakha \footnote{Samuel Reynolds Pritchard Professor of Engineering, Center for Sustainable Mobility, Virginia Tech Transportation Institute, 3500 Transportation Research Plaza (0536)
Blacksburg, VA 24061, USA. Corresponding author (hrakha@vt.edu).}}

\begin{abstract}
In this paper, empirical data from the literature are used to develop general power models that capture the impact of a vehicle position, in a platoon of homogeneous vehicles, and the distance gap to its lead (and following) vehicle on its drag coefficient. These models are developed for light duty vehicles, buses, and heavy duty trucks. The models were fit using a constrained optimization framework to fit a general power function using either direct drag force or fuel measurements. The model is then used to extrapolate the empirical measurements to a wide range of vehicle distance gaps within a platoon. Using these models we estimate the potential fuel reduction associated with homogeneous platoons of light duty vehicles, buses, and heavy duty trucks. The results show a significant reduction in the vehicle fuel consumption when compared with those based on a constant drag coefficient assumption. Specifically, considering a minimum time gap between vehicles of $0.5 \; secs$ (which is typical considering state-of-practice communication and mechanical system latencies) running at a speed of $100 \; km/hr$, the optimum fuel reduction that is achieved is $4.5 \%$, $15.5 \%$, and $7.0 \%$ for light duty vehicle, bus, and heavy duty truck platoons, respectively. For longer time gaps, the bus and heavy duty truck platoons still produce fuel reductions in the order of $9.0 \%$ and $4.5 \%$, whereas light duty vehicles produce negligible fuel savings.
\end{abstract}


\end{frontmatter}

\section{Introduction}\label{Introduction}
The \textbf{objectives} of this paper are two-fold. First, we develop general power models that capture the impact of a vehicle position, in a platoon of homogeneous vehicles, and the distance gap to its lead (and following) vehicle on its drag coefficient. These models are developed for light duty vehicles, buses, and heavy duty trucks. Second, we use these models to estimate the potential fuel reduction associated with homogeneous platoons of light duty vehicles, buses, and heavy duty trucks.

The \textbf{contributions} of the paper is that it is the first effort to develop analytical models that relate the vehicle's drag coefficient to a vehicle's position, in a platoon of homogeneous vehicles, and the distance gap to its lead (and following) vehicle on its drag coefficient.

\subsection{Literature Review and Background}
Platooning is gaining momentum as an efficient approach to increase roadway capacity and reduce vehicle fuel consumption, as several studies have suggested \cite{bonnet2000fuel,bonnet2000fuelchauffeur,al2010experimental,tsugawa2011automated,liang2016heavy,liang2016fuel,humphreys2016evaluation,tsugawa2016review,gnatowska2018influence}. One of the key factors behind this reduction in fuel consumption is the relationship between the inter-platoon distance gap and the drag forces. The drag force generated on a vehicle consists of two main components, namely: (i) the skin friction drag,  and (ii) the form drag.
The skin friction drag depends mainly on the roughness and the total area of the vehicle subjected to the air flow. This type of drag is not affected by the distance gap between vehicles. 
The most important type of drag that is affected by driving in a platoon/convoy is the form drag. The form drag is a function of the vehicle shape and flow around it. This type of drag is dependent on how quickly and smoothly the air that separates from the vehicle rejoins downstream of the vehicle, i.e. wake shape and turbulence level. In other words, the more the vehicle shape is streamlined, the less the form drag is. This type of drag can benefit the following vehicle by reducing its frontal dynamic pressure when following another vehicle at a closer spacing. This effect is observed in nature where birds fly in a streamline/wake of each other \cite{hummel1995formation,weimerskirch2001energy} known as slip-streaming or drafting and was mimicked in fighter aircraft \cite{allen2002string}.
Hence having two vehicles (one ahead and another behind) driving at a close distance gap affects the pressure forces on the vehicle, thus reducing the aerodynamic resistance force and producing fuel savings. However, the effect at a very close distance gap depends on some geometrical aspects and the type of vehicle platoons \cite{bonnet2000fuel,lammert2014effect,smith2014aerodynamic,humphreys2016evaluation,gheyssens2016effect,tsugawa2013final}, i.e. light duty vehicles (LDVs), buses or heavy duty trucks (HDTs). In other words, the effect of drag forces at very close spacings encounter an adverse behavior.

Experimental work done by Zabat et al. \cite{zabat1995aerodynamic} on LDV platoons was used in this study. The experiment was performed on $1/8$ of the full scale model of a 1991 General Motors Lumina APV in a wind-tunnel environment with drag measurements up to distance gap of  $3$ and $2$ vehicle length for the two and three-LDV platoon respectively. The results showed a drag reduction of up to $15 \%$ for the lead vehicle and up to $30 \%$ for the trail vehicle in a two-LDV platoon at a distance gap of $0.5$ of a vehicle length. For distance gaps less than $0.5$ of a vehicle length, this effect was reversed and the lead vehicle produced a higher reduction in the drag coefficient compared to the trail vehicle. Hong et al. \cite{hong1998drag} verified this behavior at close distance gaps by performing a full-scale road test, and it was also observed in part of the wind tunnel test done by Marcu and Browand \cite{marcu1999aerodynamic} in crosswind conditions.

For the bus platoons, an experimental study documented in Ref.\cite{gotz1978aerodynamik,hucho2013aerodynamics} was performed on $1:20$ scale of a cylindrical bus-shaped bodies (equivalent to Mercedes-Benz S 80 model) in a wind-tunnel environment with drag measurements up to distance gap of $5$ bus length.
The results show a drag reduction of up to $10\%$ for the lead bus and up to $60\%$ for the second bus in a two-bus platoon at a $10 m$ distance gap. 
For HDT platoons,  most of the available data were fuel measurements for different inter-platoon distance gaps \cite{browand2004fuel,tsugawa2011automated,liang2016heavy,liang2016fuel,mcauliffe2017fuel} on a full-scale truck in a road test environment with fuel measurements up to distance gap of $2$ truck length. To compute the equivalent drag coefficient, one may use the fuel model developed in Ref. \cite{rakha2011virginia} to relate the fuel consumption to the drag forces.  In addition, one of the other sources in the literature \cite{tsugawa2013final} has the fuel measurements resulting from road test for empty trucks at very close spacing of $5 -20 \; m$ which is equivalent to time gap of $0.23 - 0.9 \; secs$. As we will show later, we are not interested in these very close spacings since they are not realistic for implementation and the fuel savings for the trucks encounter a reverse behavior as mentioned earlier. The same behavior has been reported in the wind tunnel drag measurements of Ref.\cite{zabat1994drag,zabat1995aerodynamic} and been pointed out in different sources \cite{bonnet2000fuel,lammert2014effect,smith2014aerodynamic,humphreys2016evaluation,gheyssens2016effect}. However, we consider all the data \cite{tsugawa2013final,roberts2016confidence} to validate the model for the two-HDT platoon.
In general, the dependence of the drag coefficient on the inter-platoon distance gap acts in favor of  reduction of the resistance forces but may add complexity to the platoon car-following controller design \cite{gong2016constrained,zheng2017distributed} through the non-linearity introduced by coupling the vehicle-platoon model, i.e. the drag coefficient is now dependant on the distance gap between vehicles in the platoon, $C_D  = f(G = x_i - x_{i-1})$. The accurate modeling of the drag interaction between vehicles makes the controller design more efficient when it comes to finding the optimal control action using either robust or model predictive techniques \cite{richards2007robust,calafiore2012robust,li2013robust,zheng2017distributed} and reduces the uncertainty in the model \cite{mayne2014model}. In other words, the modeling of the drag coefficient improves the efficiency and accuracy of the optimization problem and in turns improves the control action needed as mentioned in Ref.\cite{zheng2017distributed}. In addition, for optimization problems that explicitly optimize fuel savings, modeling the fuel consumption accurately requires an analytic relationship between the drag coefficient and the platoon distance gap. Finally, modeling the impact of platooning on the drag coefficient is critical to quantifying the fuel/energy consumption impacts of platooning strategies. Furthermore, in quantifying the fuel reductions associated with vehicle platooning strategies, the drag coefficient of all vehicles in a platoon for the full range of distance gaps is needed, which is not available from measurement/numerical data. Hence, the need for an analytic function that describes the relation between the drag coefficient and the inter-platoon distance gap to extrapolate the data beyond the measurement/numerical spectrum is inevitable.
\subsection{Paper Contribution and Layout}
The two main contributions of this paper are: (1) we develop and present a unified model that characterizes the impact of the inter-vehicle distance gap and position in a platoon on the vehicle's drag coefficient; and (2) we use this model to quantify the energy/fuel savings associated with homogeneous platoons of LDVs, buses, and HDTs. Specifically, this developed drag model is used to provide an analytic function that describes the relation between the drag coefficient and inter-platoon distance gap: (i) quantify the potential fuel consumption savings for different homogeneous platoons at a wider range of distance gaps beyond existing  empirical measurements, (ii) quantify the potential fuel consumption savings for different homogeneous platoons for a new vehicle type without the need to perform an experimental/numerical study, (iii) to be used when the fuel consumption in the objective to be minimized \cite{li2012minimum} and for designing the controller associated with this objective \cite{li2012minimum} or other ones, i.e. maintaining time-headway for stability \cite{bian2018reducing} or minimizing the error for vehicle following control \cite{zheng2017distributed}.  

In this paper, we examine the effect of the inter-vehicle platoon distance gap on the potential of fuel reduction for LDV, bus, and HDT platoons. 
The outline of the paper is as follows. In Section \ref{Drag and Fuel Measurements for Car Bus, and Truck Platoons}, we present the empirical data available for each type of platoon. For LDV platoons \cite{zabat1994drag,zabat1995drag,zabat1995aerodynamic}, the available data are the drag measurements on a $1:8$ vehicle in wind tunnel testing. For bus platoons \cite{hucho2013aerodynamics}, the available data are drag measurements through wind tunnel tests on $1:20$ scale cylindrical bus model \cite{gotz1978aerodynamik}. For the HDT platoon modeling \cite{browand2004fuel,mcauliffe2017fuel}, the data available for the two- and three-HDT platoons is  fuel measurements through full-scale road testing. The fuel data for the truck platoon is used to compute the equivalent drag coefficient using the fuel model developed in Ref.\cite{rakha2011virginia}. 
In Section \ref{Fitting Function Drag Measurements for Car, Bus, and Truck Platoons}, we present the optimization framework used to fit the data for the drag coefficient and to extend the data for a range of distance gaps beyond that in the wind tunnel and road tests. 
In Section \ref{Fuel Curves for Car, Bus, and Truck Platoons}, we investigate the effect of the drag coefficient function on the potential fuel reduction for different vehicle types. In addition, we validate the two-HDT platoon model using the CFD \cite{humphreys2016evaluation} and fuel data \cite{roberts2016confidence}.
In Section \ref{Conclusion}, we summarize the results and discuss the impact of the current work.
\section{Drag and Fuel Measurements for LDV, Bus, and HDT Platoons}
\label{Drag and Fuel Measurements for Car Bus, and Truck Platoons}
\begin{figure}[H] 
\begin{center}
  \includegraphics[scale=0.6]{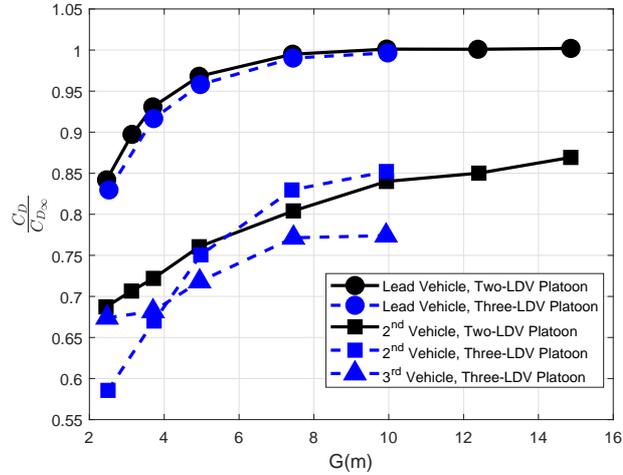}
  \caption{Drag Coefficient ratio for empirical data, $C_D/C_{D_\infty}$, for each vehicle in two- and three-LDV platoons versus distance gap, $G$, between vehicles from the experimental work done in Ref.\cite{zabat1995aerodynamic}. The drag coefficient is normalized relative drag coefficient of a single vehicle, i.e. $C_{D_\infty}$. }\label{CD_Data_Car_Platoon}
\end{center}
\end{figure}

\begin{figure}[H] 
\begin{center}
  \includegraphics[scale=0.6]{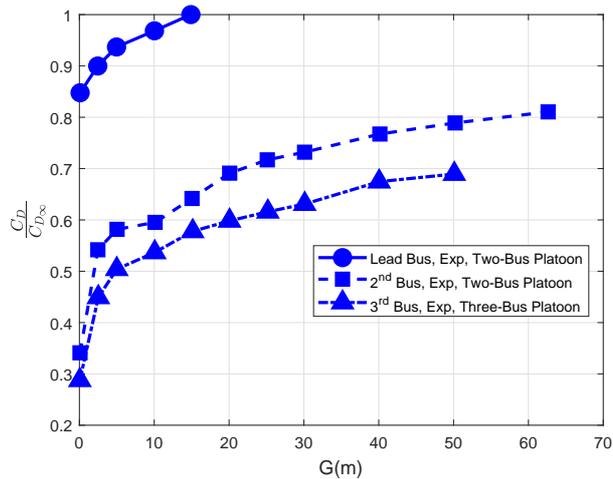}
  \caption{Drag Coefficient ratio for empirical data, $C_D/C_{D_\infty}$, for each bus in two- and three-bus platoons versus the distance gap, $G$, between buses from the experimental work done in Ref.\cite{hucho2013aerodynamics}. The drag coefficient is normalized by the drag coefficient of a single bus, i.e. $C_{D_\infty}$. }\label{CD_Data_Bus_Platoon}
\end{center}
\end{figure}

\begin{figure}[H] 
\begin{subfigmatrix}{2}
\subfigure [Fuel reduction ratio, $(F-F_\infty) / F_\infty$, for two-HDT platoons from Ref.\cite{browand2004fuel}. The fuel consumption is normalized with respect to a single truck fuel consumption, i.e. $F_\infty$. \label{Fuel_Data_Two_Truck_Platoon}]
{\includegraphics[scale=0.2]{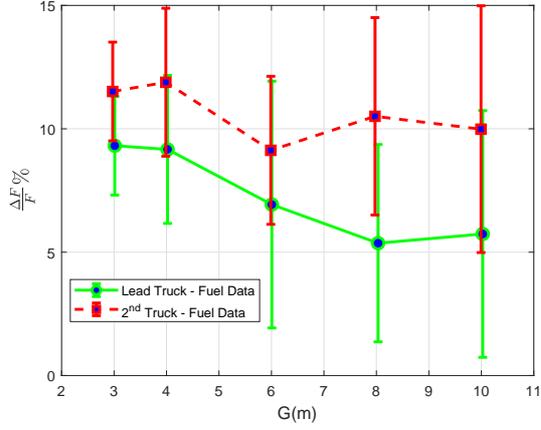}}
\subfigure [Fuel reduction ratio, $(F-F_\infty) / F_\infty$, for three-HDT platoons from Ref.\cite{mcauliffe2017fuel}. The fuel consumption is normalized with respect to a single truck fuel consumption, i.e. $F_\infty$. \label{Fuel_Data_Three_Truck_Platoon}]
{\includegraphics[scale=0.2]{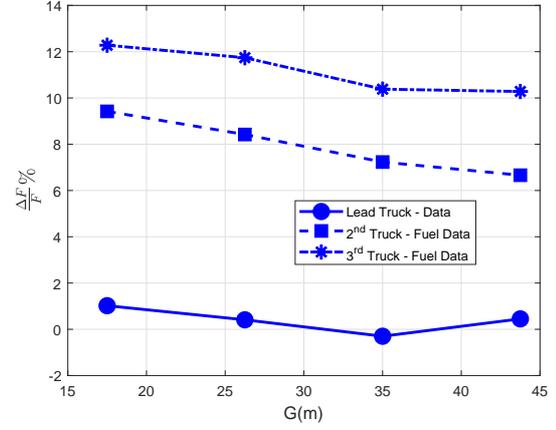}}
\subfigure[Empirical data Drag Coefficient ratio, $C_D/C_{D_\infty}$, for each vehicle in a two-HDT platoon from the CFD work done in Ref.\cite{humphreys2016evaluation}. The drag coefficient is normalized with respect to the drag coefficient of a single truck, i.e. $C_{D_\infty}$. \label{CD_Data_Truck_Platoon}]
{\includegraphics[scale=0.6]{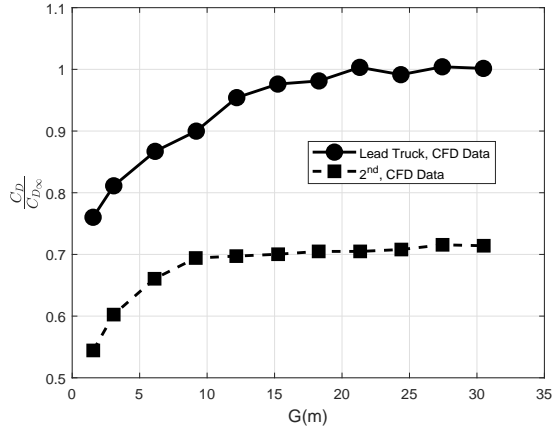}}
 \end{subfigmatrix}
\caption{Drag coefficient and fuel ratio versus distance gap, $G$, for two- and three-HDT platoons, respectively.} 
\label{Fuel_Drag_Data_Truck_Platoon}
 \end{figure}

The data for the drag measurements versus the inter-platoon distance gap for each vehicle in two- and three-LDV platoons from Refs. \cite{zabat1994drag,zabat1995drag,zabat1995aerodynamic} are shown in Figure \ref{CD_Data_Car_Platoon}. The distance gap, denoted by $G$ in all the figures and sections, is the distance measured from the rear bumper of the lead vehicle to the front bumper of the subject vehicle, i.e. for $2+$ vehicle platoons, the distance gap is symmetrical for both the front and rear of a vehicle within the platoon. For the lead vehicle measurements, the difference in the drag coefficient, $C_{D}$, for both two- and three-vehicle platoons is negligible. The lead vehicles experience a $15 \%$ reduction in the drag coefficient at a distance gap of $2.5m$. For the second vehicle measurements, the drag coefficient for the second vehicle in the three-vehicle platoon, experiences more reduction compared to the second vehicle in the two-vehicle platoon for distance gaps less than $5 m$. When the distance gap is larger than $5 m$, the behavior of the drag coefficient of the second vehicle is reversed, i.e. the drag coefficient of the second vehicle in the three-vehicle platoon experiences less reduction compared to the second vehicle in the two-vehicle platoon. Comparing the last vehicle in the three-vehicle platoon, the drag coefficient experiences more reduction over the full range of distance gaps compared to the second vehicle in the two-vehicle platoon. This is attributed to the effect of reducing the pressure on the last vehicle because of driving in the slipstream of more than one vehicle. Based on the results in Ref.\cite{zabat1995aerodynamic}, we assume that the drag reduction for the third vehicle in the platoon is almost the same as the remaining vehicles in the platoon, i.e. $C_{D}|_3 \approx C_{D}|_{3+}$. This result is also applied to Bus and HDT platoons. 
The data for the drag measurements for each bus in two- and three-bus platoons from Ref. \cite{hucho2013aerodynamics} are shown in Figure \ref{CD_Data_Bus_Platoon}. Similar to the LDV platoons, the last bus in the three-bus platoon experiences more reduction in the drag coefficient compared to the last bus in the two-bus platoon. This is attributed to the same reason discussed earlier.  
The fuel data \cite{browand2004fuel} for the two-HDT platoon is shown in Figure \ref{Fuel_Data_Two_Truck_Platoon}. The road test in done in Ref.\cite{browand2004fuel} examined a range from $3-10 m $ which is considered a small range compared to the other data considered in this work. For the three-HDT platoon \cite{mcauliffe2017fuel}, the data available also were the fuel measurements, shown in Figure \ref{Fuel_Data_Three_Truck_Platoon}. As mentioned earlier, we are not interested in this small range of distance gaps, however, we used these data as opposed to the CFD results of Ref.\cite{humphreys2016evaluation} because of their consistency with other models that are either based on wind tunnel or road test measurements. For both HDT platoons, the equivalent drag coefficients for the data in Figures [\ref{Fuel_Data_Two_Truck_Platoon},\ref{Fuel_Data_Three_Truck_Platoon}] are obtained through the fuel model developed in Ref. \cite{rakha2011virginia}, which relates the fuel consumption to the various forces via the exerted power. The procedures used to convert from fuel consumption to the drag coefficient are shown in Eqs. [\ref{Fuel_Ratio}-\ref{CD}]. 
The parameters of the vehicles used to obtain the data in the previous Figures are given in Table \ref{Vehicle_Parameters}. In the next section, we discuss the fitting procedure used to develop analytical models for LDV, bus, and HDT platoons. 
\begin{table}[H]
\centering
\caption{Vehicle characteristics used to obtain the CFD/experimental data in Figures [\ref{CD_Data_Car_Platoon}-\ref{Fuel_Drag_Data_Truck_Platoon}] for LDV, bus, and HDT platoons.}
\begin{center}
\begin{tabular}{|c|c|c|c|c|c|c|}
\cline{1-7}
\multirow{2}{*}{Vehicle Type}  & \multirow{2}{*}{Vehicle Model}  & %
    \multicolumn{5}{c|}{Parameters} \\
\cline{3-7}
& & $m (kg)$ & $length(m)$  &$width(m)$  &  $height(m)$ & $C_{D_\infty}$    \\   \cline{1-7}
\multirow{1}{*}{LDV} 
& Chevy Lumina APV & 1,700 & 4.952 & 1.877 & 1.663 & 0.367  \\ \cline{1-7} 
\multirow{1}{*}{Bus} 
& Mercedes-Benz S 80 & 16,000 & 12.000 & 2.865 & 2.865 & 0.650  \\ \cline{1-7} 
\multirow{1}{*}{HDT} 
& Volvo - VNL 670 & 29,400 & 22.710 & 2.489 & 3.353 & 0.570  \\ \cline{1-7} 
\end{tabular}
\end{center}
\label{Vehicle_Parameters}
\end{table}
\section{Fitting Drag Measurements for LDV, Bus, and HDT Platoons}
\label{Fitting Function Drag Measurements for Car, Bus, and Truck Platoons}
\begin{figure}[H] 
\begin{subfigmatrix}{2}
\subfigure[Drag Coefficient ratio, $C_D/C_{D_\infty}$, for the second vehicle in a two- LDV platoon.\label{CD_2nd_Vehicle_Optimization}]
{\includegraphics[scale=0.2]{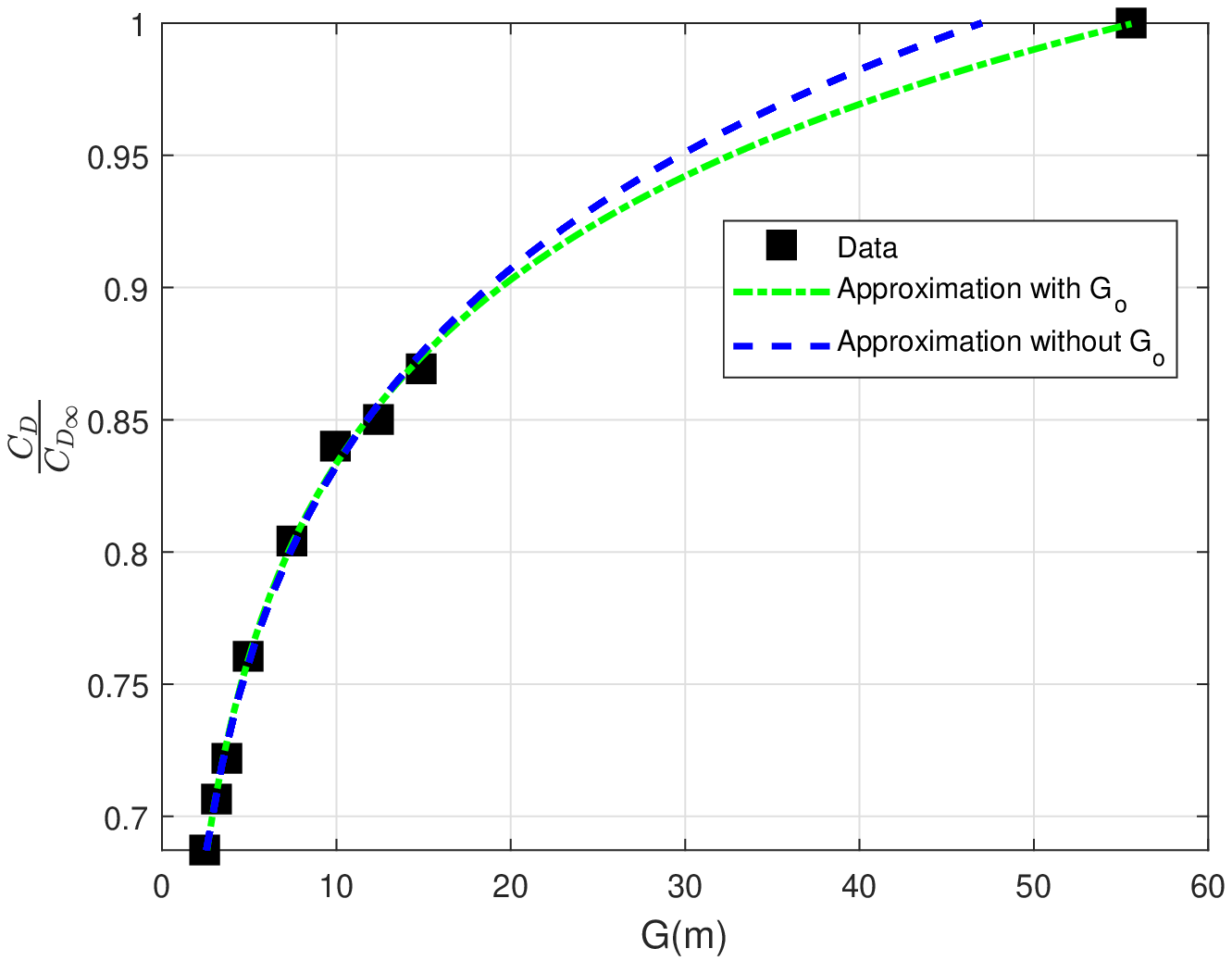}}
\subfigure [Drag Coefficient ratio, $C_D/C_{D_\infty}$, for the second truck in a two-HDT platoon.\label{CD_2nd_Truck_Optimization}]
{\includegraphics[scale=0.2]{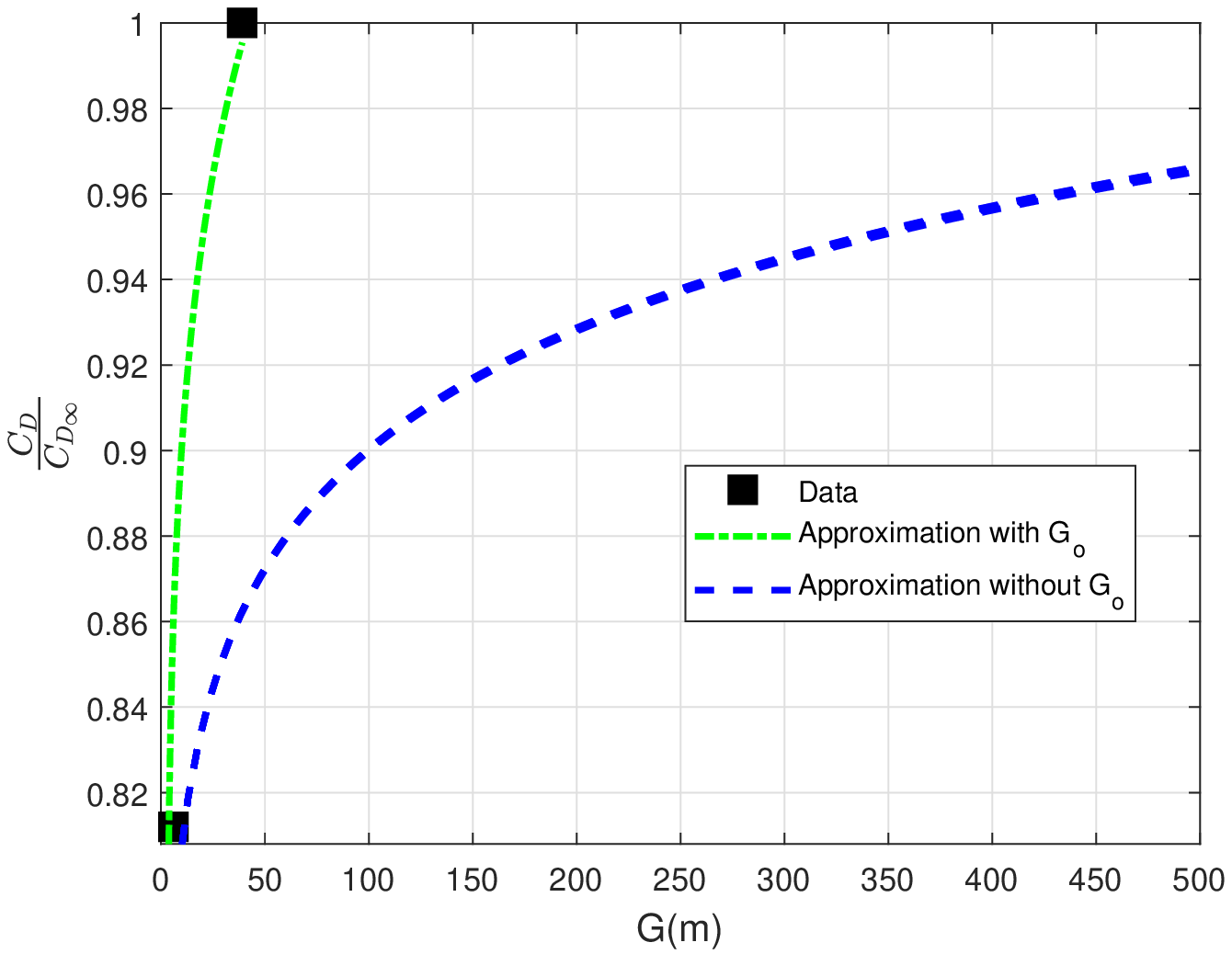}}
 \end{subfigmatrix}
\caption{Drag Coefficient ratio, $C_D/C_{D_\infty}$, for the second vehicle in a two- LDV and truck platoon versus the distance gap, $G$, for two cases where the parameter $G_o$ is excluded and included in the fitting respectively.} 
\label{CD_2nd_Optimization}
 \end{figure}

\begin{figure}[H] 
\begin{subfigmatrix}{2}
\subfigure[Drag coefficient ratio, $C_D/C_{D_\infty}$, versus distance gap for two-LDV platoon.\label{CD_Two_Car_Platoon}]
{\includegraphics[scale=0.2]{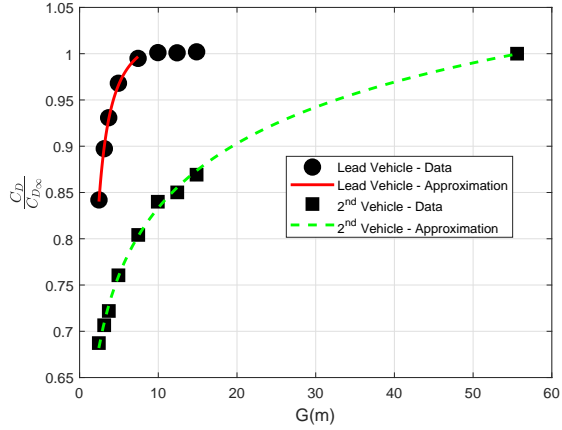}}
\subfigure [Drag coefficient ratio, $C_D/C_{D_\infty}$, versus distance gap for three-LDV platoon.\label{CD_Three_Car_Platoon}]
{\includegraphics[scale=0.2]{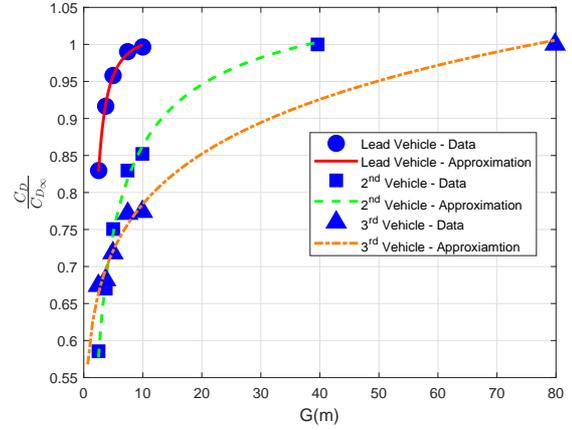}}
 \end{subfigmatrix}
\caption{Drag Coefficient ratio, $C_D/C_{D_\infty}$, for two- and three-LDV platoons. The drag coefficient is normalized with respect to the single vehicle drag coefficient, i.e. $C_{D_\infty}$.} 
\label{CD_Two_Three_Car_Platoon}
 \end{figure}

\begin{figure}[H] 
\begin{subfigmatrix}{2}
\subfigure[Drag coefficient ratio, $C_D/C_{D_\infty}$, versus distance gap for a two-bus platoon.\label{CD_Two_Bus_Platoon}]
{\includegraphics[scale=0.2]{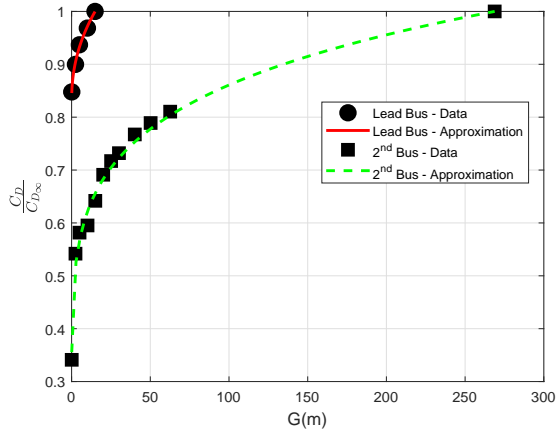}}
\subfigure [Drag coefficient ratio, $C_D/C_{D_\infty}$, versus distance gap for a three-bus platoon.\label{CD_Three_Bus_Platoon}]
{\includegraphics[scale=0.2]{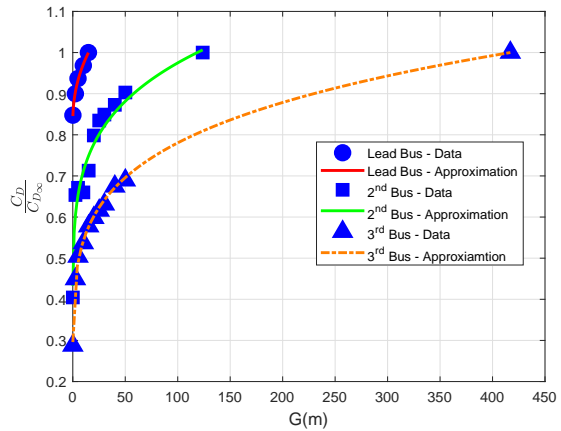}}
 \end{subfigmatrix}
\caption{Drag Coefficient ratio, $C_D/C_{D_\infty}$, for two- and three-bus platoons. The drag coefficient is normalized with the value of a single bus drag coefficient, i.e. $C_{D_\infty}$. The data in the Figure is based on the experimental measurements in Ref. \cite{hucho2013aerodynamics}.}
\label{CD_Two_Three_Bus_Platoon}
 \end{figure}

\begin{figure}[H] 
\begin{subfigmatrix}{2}
\subfigure[Drag coefficient ratio, $C_D/C_{D_\infty}$, versus distance gap for a two-HDT platoon based on the data in Ref. \cite{browand2004fuel}.\label{CD_Two_Truck_Platoon_EXP}]
{\includegraphics[scale=0.2]{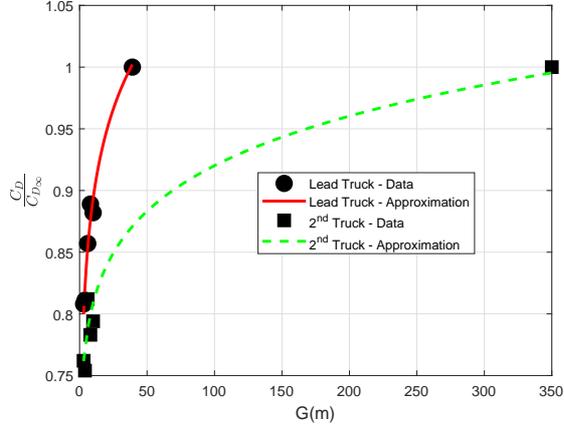}}
\subfigure [Drag coefficient ratio, $C_D/C_{D_\infty}$, versus distance gap for a three-HDT platoon based on the  data in Ref. \cite{mcauliffe2017fuel}.\label{CD_Three_Truck_Platoon_EXP}]
{\includegraphics[scale=0.2]{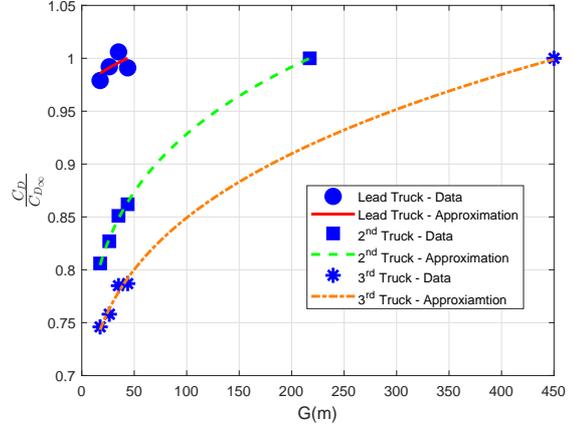}}
 \end{subfigmatrix}
\caption{Drag Coefficient ratio, $C_D/C_{D_\infty}$, for two- and three-HDT platoons. The drag coefficient is normalized with the value of a single truck drag coefficient, i.e. $C_{D_\infty}$. The data in parts \ref{CD_Two_Truck_Platoon_EXP} and \ref{CD_Three_Truck_Platoon_EXP}  is obtained by transforming the fuel measurements from Ref.\cite{browand2004fuel} and Ref. \cite{mcauliffe2017fuel} to drag data using Eqs. \eqref{Instant_Power}-\eqref{CD}.} 
\label{CD_Two_Three_Truck_Platoon}
 \end{figure}

In this section, we fit models to the empirical data for LDV, bus, and HDT platoons using a general power function as given in Eq.\eqref{General_Power_Function}. 
\begin{equation}\label{General_Power_Function}
    y = a x^b + c
\end{equation}
The advantage of using a power rather than a polynomial function is : (i) the power function varies monotonically with respect to the independent variable \cite{tadakuma2016prediction}, i.e. $y$ either continuously increases or decreases over the entire $x$ domain depending on the signs of the coefficients $a$ and $b$, and (ii) the power function has a horizontal asymptote, i.e. the distance between the power function and the horizontal axis reaches zero as $x$ approaches infinity. The second property of the power function represents an inherent property of the drag coefficient. For our case here, our objective is to have the drag function monotonically increase with the distance gap up to the drag coefficient of a single vehicle ($C_{D_\infty}$). The drag coefficient over a broad range of distance gaps can be defined as:
\begin{equation}\label{General_CD}
\frac{C_D}{C_{D_\infty}} =\left\{\begin{array}{l}
a G ^b + c, \; \; \;0 
< G \leq G_o   \\ \\
1, \; \; \;   
G \geq G_o 
\end{array}\right.\;
\end{equation}
where $G_o$ is the critical distance gap above which the drag force on a vehicle is not affected by the presence of other vehicles either in front or behind it, i.e. $C_D/C_{D_\infty} (G_o) = 1$. $C_{D_\infty}$ is the drag coefficient of a single vehicle in the absence of any other vehicles in its vicinity. The objective of the curve fitting is to find the parameters of the power function that best represent the empirical data for each case in Figures[\ref{CD_Data_Car_Platoon},\ref{CD_Data_Bus_Platoon},\ref{Fuel_Drag_Data_Truck_Platoon}], i.e. the optimal parameters that minimize the error between the function and the measurements.  The power function was used before in Ref. \cite{tadakuma2016prediction} to construct the drag coefficient reduction of a vehicle in a platoon as a function of its maximum deficit velocity rate inside the wake. Since we do not have measurements beyond certain distance gaps, the point $G_o$ is obtained by extrapolation. This is due to the nature of the underlying experiment or computational resources for each case that allowed only to span a short spectrum of the vehicle distance gap. However, for a set of initial conditions, a different local optimum solution could be obtained with different values of $G_o$ when extrapolating the curve. Hence, to determine the value of the critical distance gap, $G_o$, the $G_o$ value was assumed to be unknown and was estimated through the optimization procedure. The nonlinear least square data fitting is defined as follows 
\begin{equation}\label{NONLSQ_CD}
    \underset{\bm{z}}{\textmd{\rm min}}  \sum_{j=1}^{N_p} 
    \left( \frac{C_D}{C_{D_\infty}}(G_j)  -  \frac{C_D}{C_{D_\infty}} (G_j)\rvert_M \right)^2 
\end{equation}
subjected to the bound constraint on $G_o$
\begin{equation}\label{Constraint_So}
\begin{split}
G_{o_l} & \leq G_o \leq G_{o_u} 
\end{split}
\end{equation}
where $\bm{z} = \{a, b, c, G_o\}^T$ is the vector of the design variables, $N_p$ is the number of empirical observations available for each case, and the subscript $M$ stands for measurements.  We used the nonlinear least square function \textbf{lsqnonlin} in Matlab with either \textbf{Levenberg-Marquardt} \cite{levenberg1944method} or \textbf{trust-reflective-region} \cite{byrd1987trust} algorithms. The latter algorithm is used when a bound constraint on $G_o$ is needed. In all the cases, the nonlinear least square optimization is used without constraints on $G_o$ except for the case of the trail truck for the two- \cite{browand2004fuel} and three-HDT platoons \cite{mcauliffe2017fuel} where an unconstrained fitting gives an unreasonable value for $G_o$, i.e. $G_o{_{trail}} \approx 1000$, which is equivalent to 40 truck lengths. This is attributed to the high uncertainty in the data for the trail trucks \cite{browand2004fuel,mcauliffe2017fuel}. The constraints for these two cases are chosen such that it satisfies the relative value of  $G_o$ derived from the results of LDV and Bus platoons. The optimum parameters for each vehicle type and platoon configuration are summarized in Table \ref{Optimization_Parameters}. 
In all cases considered, the drag coefficient ratio of the lead vehicle reaches unity very quickly compared to the middle and the last vehicle, i.e. the critical distance gap at which the lead vehicle is no longer influenced by the rear ones.
\begin{table}[H]
\centering
\caption{Drag coefficient parameters for each vehicle in LDV, bus, and HDT platoon.}
\begin{center}
\begin{tabular}{|c|c|c|c|c|c|c|}
\cline{1-7}
\multirow{2}{*}{Vehicle Type}  & \multirow{2}{*}{Platoon Size} & \multirow{2}{*}{Vehicle Position} & %
    \multicolumn{4}{c|}{Parameters} \\
\cline{4-7}
 & & & $a$ & $b$ & $c$ & $G_o(m)$ \\ \cline{1-7}
\multirow{5}{*}{LDV} & \multirow{2}{*}{Two}  
& Lead & -0.7575 & -1.5225 & 1.0325 & - \\ \cline{3-7} 
& & Trail & -1.7834 & -0.0672 & 2.3614 & 55.72 \\ \cline{2-7} 
& \multirow{3}{*}{Three} 
& Lead & -0.8906 & -1.6679 & 1.0185 & - \\ \cline{3-7} 
& & Middle & -0.8985 & -0.5126 & 1.1393 & 39.62 \\ \cline{3-7} 
& & Trail & -0.5953 & -0.1197 & 1.1393 & 79.75 \\ \cline{1-7} 
\multirow{5}{*}{Bus} & \multirow{2}{*}{Two}  
& Lead & 0.0506 & 0.4527 & 0.8280 & - \\ \cline{3-7} 
& & Trail & 0.2921 & 0.1862 & 0.1724 & 268.79 \\ \cline{2-7} 
& \multirow{3}{*}{Three} 
& Lead & 0.0506 & 0.4527 & 0.8280 & - \\ \cline{3-7} 
& & Middle & 0.2622 & 0.2104 & 0.2728 & 127.68 \\ \cline{3-7} 
& & Trail & 0.2250 & 0.2159 & 0.1722 & 416.98 \\ \cline{1-7} 
\multirow{5}{*}{HDT} & \multirow{2}{*}{Two}  
& Lead & 0.7231 & 0.0919 & 0.000 & 34.0181 \\ \cline{3-7} 
& & Trail & 0.2241 & 0.1369 & 0.5016 & 320 \\ \cline{2-7} 
& \multirow{3}{*}{Three} 
& Lead & 0.0035 & 0.5997 & 0.9662 & - \\ \cline{3-7} 
& & Middle & 0.1522 & 0.2111 & 0.5260 & 217.27 \\ \cline{3-7} 
& & Trail & 0.0726 & 0.2842 & 0.5794 & 480.00 \\ \cline{1-7} 
\end{tabular}
\end{center}
\label{Optimization_Parameters}
\end{table}
To illustrate the benefits of including the $G_o$ parameter in the fitting procedure, a comparison between the two cases is conducted for the trail vehicle and truck in the two-LDV and two-HDT platoons, respectively, as illustrated in Figure \ref{CD_2nd_Vehicle_Optimization}. For the case of the LDV, where $G_o$ is not a parameter, its value is computed as $47.0 ; m$ by extrapolating the curve to the unity value of the drag coefficient ratio. For the case where $G_o$ is included in the fitting, its value is $55.7 \; m$. The residual error is $0.6387 \times 10^{-8}$ and $0.7734 \times 10^{-8}$ for the inclusion and exclusion of the $G_o$ in the optimization, respectively. Not only does the inclusion of the $G_o$ parameter in the optimization yield a better optimal solution, but it also provides a natural formalism for obtaining the value of $G_o$ which (its relative value inside the platoon) should be consistent for different platoon sizes as will be discussed in the next Figures. Similarly for the case of the trail truck in the two-HDT platoon in Figure \ref{CD_2nd_Truck_Optimization}, the inclusion of $G_o$ as a variable in the optimization yields much better results, with the residual error decreasing from $54.89 \times 10^{-8}$ to $6.5218 \times 10^{-8}$. Consequently, for the case of the trail truck in the two- and three-HDT platoons, a constraint on $G_o$ has to be enforced since performing the optimization with  $G_o$ unbounded yields unrealistic values ($G_o \approx 1000m$). The selection of constraint bounds on $G_o$ are detailed in the next section. 
In Figure \ref{CD_Two_Three_Car_Platoon}, the data fitting results are shown for two- and three-LDV platoons. The experimental data are extracted from the work of Zabat et al. \cite{zabat1994drag,zabat1995drag,zabat1995aerodynamic}. As illustrated in the Figures [\ref{CD_Two_Car_Platoon},\ref{CD_Three_Car_Platoon}], the models for the lead vehicle in both the two- and three-LDV platoons are very similar. Noteworthy is the fact that the distance gap needed for the drag coefficient ratio of the following vehicle to reach unity after the drag coefficient of its lead reaches unity are close for the two different platoons, i.e. the difference between the critical distance gap of each vehicle in two- and three-LDV platoons, $G_{o_i} - G_{o_{i-1}}$ is small.  For the two-LDV platoon, the relative distance gap is $ G_{o_{trail}} - G_{o_{lead}} \approx 45 m$. For the three-LDV platoon, the relative distances between the lead and the middle and middle and the trail are $ G_{o_{middle}} - G_{o_{lead}} \approx 30 m$ and $ G_{o_{trail}} - G_{o_{middle}} \approx 40m$, respectively. This is a natural result from the optimization, which agrees with the physical meaning of the parameter $G_o$. Not only obtaining this agreement for the value of $G_o$ would be hard using trial and error but enforcing it through the inclusion of constraints would bias the optimization. Depending on the accuracy of the measurements, the parameter $G_o$ will be closer to reality. The inclusion of the parameter $G_o$ in the optimization improves the accuracy of the extrapolation of the data to a wide range of distance gaps and is considered as a predictor for the no-influence point between vehicles in the platoon. The $G_o$ parameter should thus be considered as an upper bound for the distance gap between vehicles when designing the platoon controller \cite{gong2016constrained,zheng2017distributed} since the objective is to keep the vehicles within a platoon as close as possible to maximize their fuel savings. 
In Figure \ref{CD_Two_Three_Bus_Platoon}, the results for the two and three bus platoon are shown. The fitting is based on the experimental measurement in Figure \ref{CD_Data_Bus_Platoon} collected from Ref. \cite{hucho2013aerodynamics}. Based on the measurements for the lead car in Figure \ref{CD_Data_Car_Platoon}, the curve for the lead bus in the presence of one or more buses behind is assumed the same. Hence the data and the approximation curves in Figures [\ref{CD_Two_Bus_Platoon},\ref{CD_Three_Bus_Platoon}] are identical. The data shown in Figure \ref{CD_Three_Bus_Platoon} for the case of the second bus in a three-bus platoon is obtained by assuming that the drag coefficient for the second bus in a two-bus platoon is the average of the result of the last bus in a two- and three-bus platoon. This approximation is based on the observation of the car results in Figure \ref{CD_Data_Car_Platoon}. Similar to the LDV platoon in Figure \ref{CD_Two_Three_Car_Platoon}, the relative distance gap between the trail and the lead bus in the two-bus platoon is close to the relative distance gap between the middle and the trail bus in the three-bus platoon, i.e. $ G_{o_{trail}} - G_{o_{lead}} \approx 253 m$ and $ G_{o_{trail}} - G_{o_{middle}} \approx 285 m$.
In Figure \ref{CD_Two_Three_Truck_Platoon}, the results for two- and three-HDT platoons are shown. The drag coefficient for the two and three-HDT platoons in Figures [\ref{CD_Two_Truck_Platoon_EXP},\ref{CD_Three_Truck_Platoon_EXP}] are based on the measurements in Ref. \cite{mcauliffe2017fuel}. The drag coefficients  are calculated from the fuel measurements via the relations defined in Eqs. [\eqref{Instant_Power}-\eqref{Fuel_Consumption}]. Noteworthy is the fact that for the case of the trail truck in the two- and three-HDT platoons, the bounding constraint on $G_o$ is needed given the uncertainty in the data. The constraint was chosen to yield similar $G_o$ values to those derived for LDV and bus platoons, as illustrated in Figures [\ref{CD_Two_Three_Car_Platoon},\ref{CD_Two_Three_Bus_Platoon}]. In other words, the bounds are chosen such that the optimization yields a relative distance gaps between the trail and the lead truck in the two-HDT platoon similar to the relative distance gap between the middle and the trail truck in the three-truck platoon, i.e. $ G_{o_{trail}} - G_{o_{lead}} \approx 263 m$ and $ G_{o_{trail}} - G_{o_{middle}} \approx 287 m$.  If the constraint was not invoked, the optimization would yield a value of $G_o \approx 1000 m$ for these cases. This is the only case where constraints on $G_o$ were invoked and is based on the physical intuition concluded from the results of the LDV and bus platoons. 
\section{Fuel Curves for LDV, Bus, and HDT Platoons}
\label{Fuel Curves for Car, Bus, and Truck Platoons}
\begin{figure}[H] 
\begin{subfigmatrix}{2}
\subfigure[Fuel reduction ratio, $(F-F_\infty) / F_\infty$, versus distance gap for two-LDV platoon. \label{Fuel_Two_Car_Platoon}]
{\includegraphics[scale=0.2]{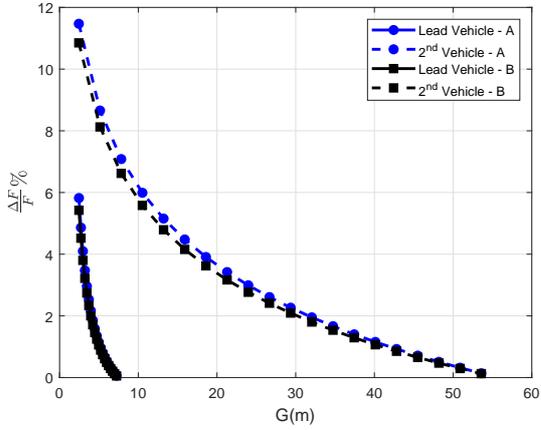}}
\subfigure [Fuel reduction ratio, $(F-F_\infty) / F_\infty$, versus distance gap for three-LDV platoon. \label{Fuel_Three_Car_Platoon}]
{\includegraphics[scale=0.2]{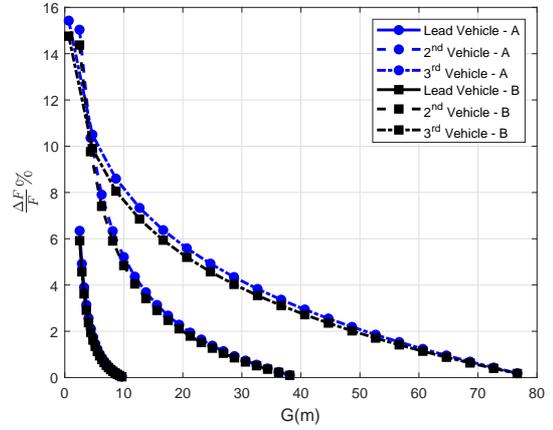}}
 \end{subfigmatrix}
\caption{Fuel reduction ratio, $(F-F_\infty) / F_\infty$, for two- and three-LDV platoons of type $A$ and $B$. The fuel consumption is normalized with respect to a single vehicle fuel consumption, i.e. $F_\infty$.} 
\label{Fuel_Two_Three_Car_Platoon}
 \end{figure}

\begin{figure}[H] 
\begin{subfigmatrix}{2}
\subfigure[Fuel reduction ratio, $(F-F_\infty) / F_\infty$, versus distance gap for two-bus platoon. \label{Fuel_Two_Bus_Platoon}]
{\includegraphics[scale=0.2]{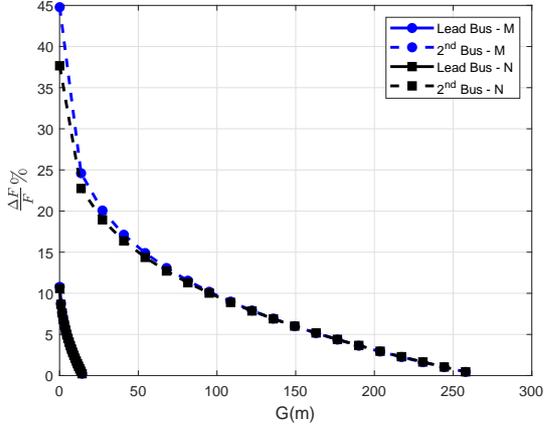}}
\subfigure [Fuel reduction ratio, $(F-F_\infty) / F_\infty$, versus distance gap for three-bus platoon. \label{Fuel_Three_Bus_Platoon}]
{\includegraphics[scale=0.2]{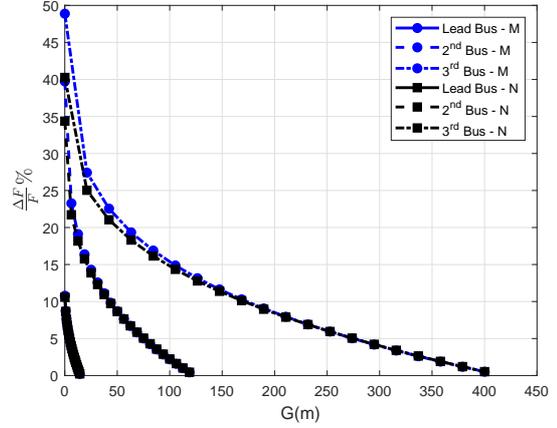}}
 \end{subfigmatrix}
\caption{Fuel reduction ratio, $(F-F_\infty) / F_\infty$, for two- and three-bus platoons of bus types $M$ and $N$. The fuel consumption is normalized with respect to the value of a single bus fuel consumption, i.e. $F_\infty$.} 
\label{Fuel_Two_Three_Bus_Platoon}
 \end{figure}

\begin{figure}[H] 
\begin{subfigmatrix}{2}
\subfigure[Fuel reduction ratio, $(F-F_\infty) / F_\infty$, versus distance gap for two-HDT platoon. The fuel data is from Ref.\cite{browand2004fuel}. \label{Fuel_Two_Truck_Platoon}]
{\includegraphics[scale=0.2]{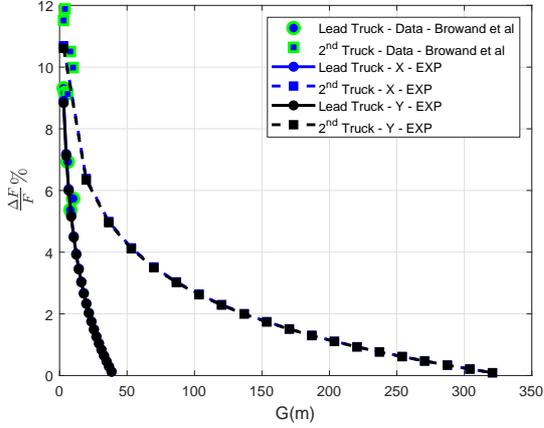}}
\subfigure [Fuel reduction ratio, $(F-F_\infty) / F_\infty$, versus distance gap for three-HDT platoon.  The fuel data is from Ref.\cite{mcauliffe2017fuel}. \label{Fuel_Three_Truck_Platoon}]
{\includegraphics[scale=0.2]{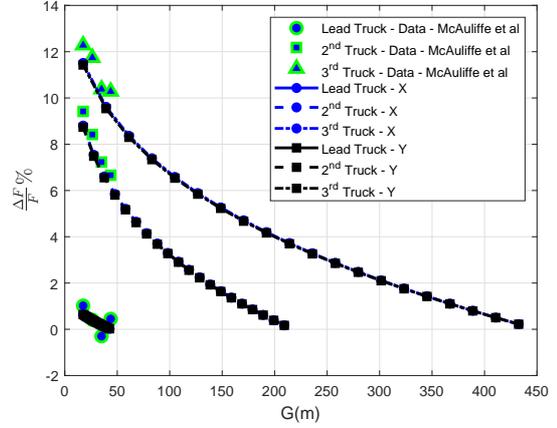}}
 \end{subfigmatrix}
\caption{Fuel reduction ratio, $(F-F_\infty) / F_\infty$, for two- and three-HDT platoons of type $X$ and $Y$. The fuel consumption is normalized with the value of a single truck fuel consumption, i.e. $F_\infty$.} 
\label{Fuel_Two_Three_Truck_Platoon}
 \end{figure}

\begin{figure}[H] 
\begin{subfigmatrix}{2}
\subfigure[Average fuel reduction ratio, $(F-F_\infty) / F_\infty$, for three-LDV platoon and different speeds as a function of time gap for type A vehicle.\label{Fuel_Velocity_GapTime_Three_Car_Platoon}]
{\includegraphics[scale=0.2]{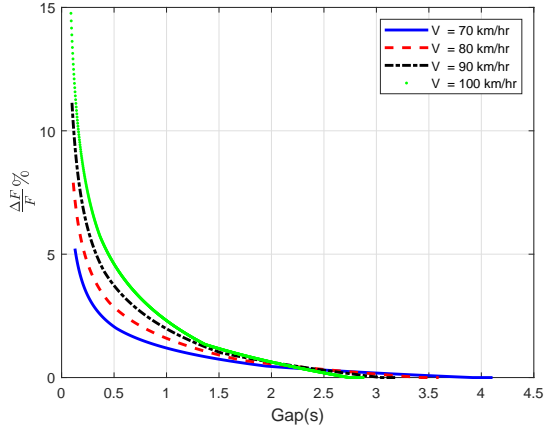}}
\subfigure[Average fuel reduction ratio, $(F-F_\infty) / F_\infty$, for three bus platoons and different speeds as a function of time gap for type M bus.\label{Fuel_Velocity_GapTime_Three_Bus_Platoon}]
{\includegraphics[scale=0.2]{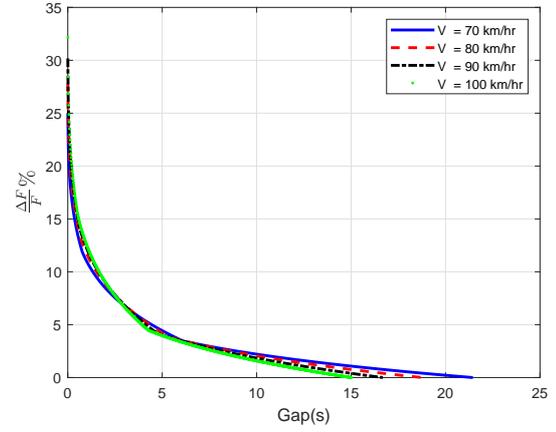}}
\subfigure[Average fuel reduction ratio, $(F-F_\infty) / F_\infty$, for three-HDT platoons and different speeds as a function of time gap for type X truck.\label{Fuel_Velocity_GapTime_Three_Truck_Platoon}]
{\includegraphics[scale=0.2]{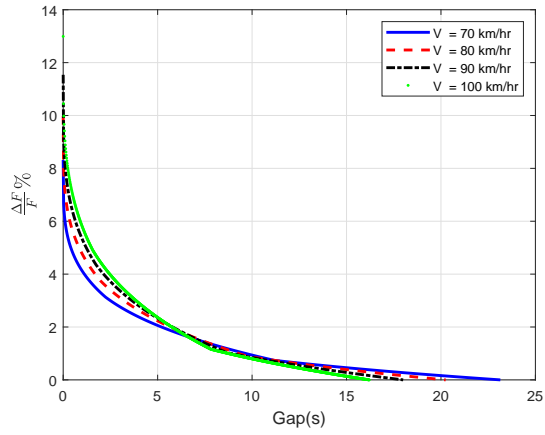}}
 \end{subfigmatrix}
\caption{Average fuel reduction ratio for different platoon and different speeds as a function of the time gap.} 
\label{Fuel_Velocity_GapTime_Platoons}
 \end{figure}

\begin{figure}[H] 
\begin{center}
  \includegraphics[scale=0.6]{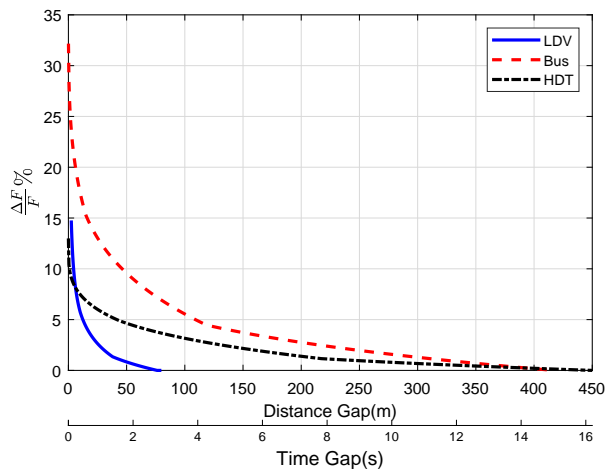}
  \caption{Average fuel reduction ratio, $(F-F_\infty) / F_\infty$, for three vehicle platoons as a function of distance gap/time gap at speed of $V = 100 \; km/hr$.}\label{Fuel_distance gap_GapTime_Platoons}
\end{center}
\end{figure}

\begin{figure}[H] 
\begin{subfigmatrix}{2}
\subfigure[Fuel reduction ratio, $(F-F_\infty) / F_\infty$, for the lead truck as a function of distance gap.\label{Fuel_Two_Truck_Validation_Confidence_Report_Lead}]
{\includegraphics[scale=0.2]{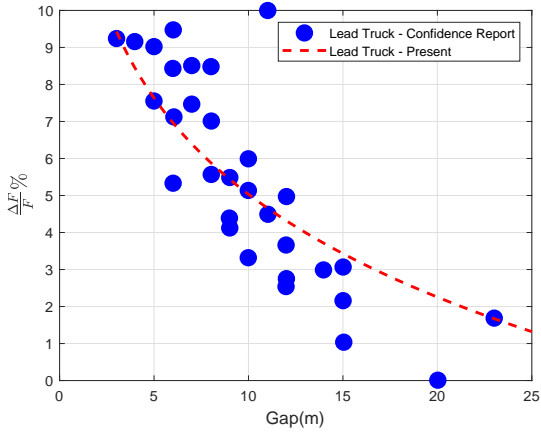}}
\subfigure[Fuel reduction ratio, $(F-F_\infty) / F_\infty$, for the trail truck as a function of distance gap.\label{Fuel_Two_Truck_Validation_Confidence_Report_Trail}]
{\includegraphics[scale=0.2]{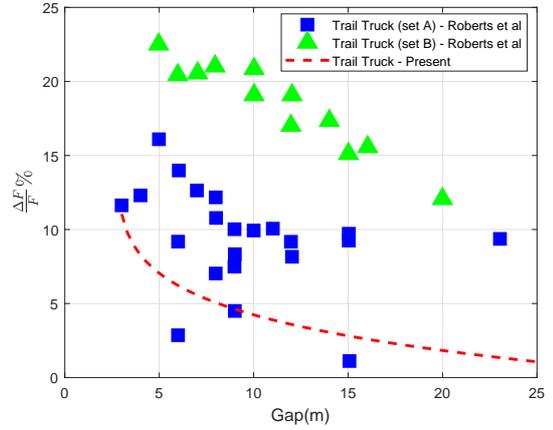}}
 \end{subfigmatrix}
\caption{Fuel reduction ratio, $(F-F_\infty) / F_\infty$, versus distance gap for two-HDT platoon. The curves represent the proposed model and the data are from the confidence report on two-HDT platooning \cite{roberts2016confidence}.} 
\label{Fuel_Two_Truck_Validation_Confidence_Report}
 \end{figure}

In this Section, we examine the effect of using the drag model developed in Section \ref{Fitting Function Drag Measurements for Car, Bus, and Truck Platoons} on the platoon fuel consumption. The fuel model is developed by Rakha et al. \cite{rakha2011virginia} to calculate the instantaneous fuel consumption. The instantaneous power in $kW$ is calculated as 
\begin{equation}\label{Instant_Power}
    P(t) = \left(\frac{R(t) + 1.04 m a(t)}{3600 \eta_d} \right) v(t)
\end{equation}
where $m$ is the vehicle mass in $kg$, $a(t)$ is vehicle acceleration in $m/s^2$ at instant $t$, $v(t)$ is the vehicle speed in $km/h$ at instant $t$, $\eta_d$ is the drive-line efficiency, and $R(t)$ is the resistance force in $N$ at instant $t$. The resistance force $R(t)$ is calculated as 
\begin{equation}\label{Resistance_Force}
    R(t) = \frac{\rho}{25.92} C_d C_h A_f v(t)^2 + g m \frac{C_r}{1000}\left( C_1 v(t) + C_2 \right) + g m G(t)
\end{equation}
where $\rho$ is the air density at sea level and $15^o c$, $C_d$ is the vehicle drag coefficient, $C_h$ is the correction factor of elevation, $A_f$ is the vehicle frontal area in $m^2$, $g$ is the gravitational acceleration, $G(t)$ is the roadway grade, and $C_r$, $C_1$, and $C_2$ are the rolling resistance parameters. The fuel consumption is then calculated using power computed using Eq. \eqref{Instant_Power} as 
\begin{equation}\label{Fuel_Consumption}
F(t) =\left\{\begin{array}{l}
\alpha_0 + \alpha_1 P(t) + \alpha_2 P(t)^2, \; \;   P(t) \geq 0   \\ \\
\alpha_0, \; \; \;   P(t) < 0 
\end{array}\right.\;
\end{equation}
where the coefficients $\alpha_0$, $\alpha_1$, and $\alpha_2$ are calculated using the power and fuel consumed using the Environmental Protection Agency (EPA) fuel ratings \cite{rakha2011virginia}.
To obtain the equivalent drag coefficient for trucks that is shown in Figures [\ref{CD_Two_Truck_Platoon_EXP},\ref{CD_Three_Truck_Platoon_EXP}] from the fuel measurements in Figures [\ref{Fuel_Data_Two_Truck_Platoon},\ref{Fuel_Data_Three_Truck_Platoon}], the fuel ratio is defined as 
\begin{equation}\label{Fuel_Ratio}
\begin{split}
        \frac{F_\infty - F}{F_\infty} & = \Delta \\
     \Rightarrow \ 
    F & = F_\infty(1-\Delta)  \\
    & = \left( \alpha_0 + \alpha_1 P(t) + \alpha_2 P(t)^2 \right) n
\end{split}
\end{equation}
where $F_\infty$ is the fuel consumption rate of the vehicle when no other vehicles are present either in-front or behind it, $n$ is the amount of fuel consumed for the same condition.
The power is then calculated from Eq. \eqref{Fuel_Consumption} as
\begin{equation}\label{Power}
    P = \frac{-n \alpha_1 + \sqrt{n^2 \alpha_1^2-4 n \alpha_2(n\alpha_0-F)}}{2 n \alpha_2}
\end{equation}
Hence the drag coefficient from the force relation in Eq. \eqref{Resistance_Force} as
\begin{equation}\label{CD}
    C_D = \frac{ \frac{P \times 3600 \eta}{v} - R_\infty   }
    { \frac{\rho}{25.92} A_f C_h v^2 }
\end{equation}
where $R_\infty$ is the resistance force of the vehicle at no other vehicles present either in-front or in rear.
\begin{table}[H]
\centering
\caption{Parameters required for each vehicle in LDV, bus, and HDT platoon for the fuel model defined in Eq. \eqref{Fuel_Consumption}.}
\begin{center}
\begin{tabular}{|c|c|c|c|c|c|c|c|c|}
\cline{1-9}
\multirow{2}{*}{Vehicle Type}  & \multirow{2}{*}{Vehicle Model}  & %
    \multicolumn{7}{c|}{Parameters} \\
\cline{3-9}
& & $m (kg)$ & $\eta_d$  &$C_{D_\infty}$  &  $A_f (m^2)$  & $\alpha_0$ & $\alpha_1$ & $\alpha_2$ \\   \cline{1-9}
\multirow{2}{*}{LDV} 
& A & 1469 & 0.80 & 0.325 & 2.30 & 6.00e-4 & 1.90e-5 & 1.00e-6 \\ \cline{2-9} 
& B & 1550 & 0.80 & 0.24 & 2.20 & 5.00e-4 & 4.41e-5 & 1.00e-6 \\ \cline{1-9}
\multirow{2}{*}{Bus} 
& M & 8505 & 0.95 & 0.80 & 7.59 & 1.33e-3 & 6.33e-5 & 1.00e-8  \\ \cline{2-9} 
& N & 13486 & 0.95 & 0.80 & 7.38 & 8.31e-4 & 1.90e-5 & 5.34e-7 \\ \cline{1-9}
\multirow{3}{*}{HDT} 
& X & 7239 & 0.88 & 0.78 & 8.90 & 1.56e-3 & 8.10e-5 & 1.00e-8 \\ \cline{2-9} 
& Z & 12864 & 0.88 & 0.78 & 8.80 & 1.66e-3 & 8.60e-5 & 1.00e-8 \\ \cline{2-9}
& McAuliffe et al. & 8500 & 0.94 & 0.57 & 10.70 & 1.56e-3 & 8.10e-5 & 1.00e-8 \\ \cline{1-9} 
\end{tabular}
\end{center}
\label{Fuel_Vehicle_Parameters}
\end{table}
In Figure \ref{Fuel_Two_Three_Car_Platoon}, the fuel reduction ratio is shown for the case of two and three-car platoons for two different car types: A and B. The parameters for the two type of cars are given in Table \ref{Fuel_Vehicle_Parameters}. For the two-vehicle platoon in Figure \ref{Fuel_Two_Car_Platoon}, the second vehicle experiences fuel reductions more than the lead vehicle; up to $6 \%$ for the second one with no savings for the lead one at a distance gap of $10 \; m $. Similarly for the three-car platoon in Figure \ref{Fuel_Three_Car_Platoon}, the third, the second, and the lead experience up to an $8 \% $, $ 5 \%$, and $0 \%$ fuel reduction, respectively. The different car parameters have a negligible effect on the fuel reduction curves. It should be noted that these results agree with what is reported in Ref.\cite{michaelian2001quantifying} with regards to fuel savings derived from wind tunnel measurements. Specifically, they suggest that other parameters should be included in wind tunnel-based models (e.g. the turbulence level), when comparing to fuel estimates from on-road tests.
In Figure \ref{Fuel_Two_Three_Bus_Platoon}, the fuel reduction ratio is shown for the case of two- and three-bus platoons for two different bus types: M and N. The similar trend of fuel reduction in car platoons is observed here. For the two-bus platoon in Figure \ref{Fuel_Two_Bus_Platoon}, the second bus experiences fuel reductions more than the lead bus with up to $15 \%$ reductions for the second one with no savings for the lead one at a distance gap of $50 \; m $. Similarly for the three-bus platoon in Figure \ref{Fuel_Three_Bus_Platoon}, the third, the second, and the lead experience up to $20 \% $, $ 10 \%$, and $0 \%$ fuel reductions, respectively. The maximum payload used for the buses in these figures are $5000 \; kg $ and $2500 \; kg$ for type M and N, respectively. This could be found on the manufacturer website.
In Figure \ref{Fuel_Two_Three_Truck_Platoon}, the fuel reduction ratio is shown for the case of two- and three-truck platoons for two different truck types: X and Y. The fuel reduction curves follow the same behavior in the car and bus platoons. For the two-truck platoon in Figure \ref{Fuel_Two_Truck_Platoon}, the second truck experiences fuel reductions more than the lead truck by up to $8 \%$ at a distance gap of $50 \; m $. The lead one experience $0 \%$ at this distance gap using both of the two drag models. Similarly for the three truck platoon in Figure \ref{Fuel_Three_Truck_Platoon}, the third, the second, and the lead experience up to $9 \% $, $ 6 \%$, and $0 \%$ respectively at distance gap of $50 \; m$. We used the equivalent payload from Ref. \cite{mcauliffe2017fuel} to make sure the different truck are having the same weight, i.e. $W_{payload} = 22161 \; kg$ for truck X, and $W_{payload} = 16536 \; kg$ for truck Y. The weight of the truck used in the road test experiment in Ref. \cite{mcauliffe2017fuel} is $29400 \; kg$. 
In Figure \ref{Fuel_Velocity_GapTime_Platoons}, the fuel reduction ratio is shown for the case LDV, bus, and HDT platoon as a function of time gap for different vehicle speeds. Since it is more appropriate to specify gap between vehicles in terms of time (headway) in process of  the  vehicle platoon controller design \cite{gong2016constrained,zheng2017distributed,bian2018reducing,ding2018cooperative} , we transformed the Figures in terms of the time gap between vehicles. Accounting for communication, controller and mechanical latency, the minimum time gap that could be achieved will be lower bounded as $ Gap \geq 0.5 \; secs$ (see Refs. \cite{almannaa2017reducing,almannaa2019field}. Hence the optimum fuel reduction that could be attained is $4.5 \%$, $15.5 \%$, and $7 \%$ for LDV, bus, and HDT platoon respectively running at speed of $100 \; km/hr$. These results agree with what have been found in the numerical simulation by Alam et al \cite{al2010experimental} for  the HDT platoon. In addition to the fuel reductions, the lower time gap has the benefit of increasing the roadway capacity. For instance, based on the vehicle lengths in Table \ref{Vehicle_Parameters} and a travelling speed of $100 \; km/hr$, the headway of the LDV, bus, and HDT is $0.678 secs$,  $0.932 secs$, and $1.317 secs$, respectively. These are equivalent to saturation flow rates of $5,309$, $3,862$, and $2,733 \; veh/hr/lane$, respectively. These values provide significant improvements over typical base LDV saturation flow rates of $2,450 \; veh/hr/lane$ (Highway Capacity Manual).
To elaborate more on the fuel savings between different platoons, the fuel reduction for the three different type of platoons is shown in Figure \ref{Fuel_distance gap_GapTime_Platoons} as a function of distance gap/time gap at velocity of $100 \; km/hr$. The distance gap and time gap are provided on the same $x$ axis for clarity. As mentioned earlier, to account for the time delay in the vehicle mechanical response and communication between vehicles, the lower bound of the attainable time gap between vehicles is bounded by $0.5 \; secs$. As we see from Figure \ref{Fuel_distance gap_GapTime_Platoons}, the $0.5 \; secs$ is equivalent to a distance gap of $25 \; m$. This may be challenging in the case of bus and HDT platoons, where the lower bound of the time gap need to be higher than that of LDV platoons. If we go to the value of time gap of $2 \; secs$, we can see that the bus and HDT platoons still produce a significant amount of fuel reduction, with savings up to $9 \%$ and $4.5 \%$, respectively while the LDV platoons is almost $0.6 \%$.
The model for two-HDT platoon \cite{browand2004fuel} is validated by comparing its estimates to the CFD data of Ref.\cite{humphreys2016evaluation}. For the case of the lead truck, the drag coefficient model offers a very good fit with the CFD data over the full range of platoon distance gaps.  As seen from Figure \ref{CD_Data_Truck_Platoon}, the drag coefficient is asymptotic to a value less than 1.0 which is inconsistent with the measurements for either LDV, bus or HDT platoons. This finding agrees with Ref. \cite{gnatowska2018influence}, in which they concluded that the CFD data for the non-lead vehicles is inconsistent with empirical data at shorter distance gaps. Hence we did not use the CFD data for the trail truck to validate the proposed model.
\begin{figure}[H] 
\begin{center}
\includegraphics[scale=0.6]{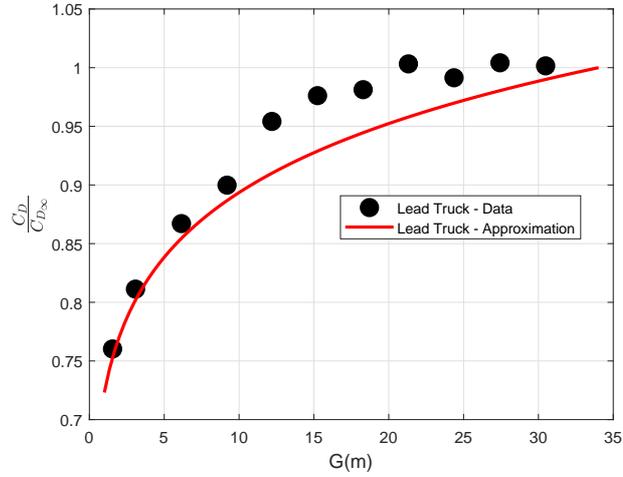}
\caption{Drag coefficient ratio, $C_D / C_{D_\infty}$, versus distance gap for two-HDT platoon. The proposed model versus the CFD results from Ref.\cite{humphreys2016evaluation}.}\label{Fuel_Two_Truck_Validation}
\end{center}
\end{figure}
In Figure \ref{Fuel_Two_Truck_Validation_Confidence_Report}, the model estimates for the two-HDT platoon are compared to the experimental measurements provided in Ref.\cite{roberts2016confidence}. As is evident from  Figure \ref{Fuel_Two_Truck_Validation_Confidence_Report_Lead}, the developed model provides a very good agreement with all the empirical data for the lead truck. On the contrary, the model estimates for the trail truck is consistent with part of the data (set A), as shown in Figure \ref{Fuel_Two_Truck_Validation_Confidence_Report_Trail}. The deviation in the other data set (set B) in Figure \ref{Fuel_Two_Truck_Validation_Confidence_Report_Trail} is attributed to the adverse behavior at low distance gaps discussed earlier which depends on different factors. e.g. the geometry of the truck.
\section{Conclusion}\label{Conclusion}
In this paper, the effect of inter-platoon position and distance gap strategy on the vehicle drag coefficient and fuel consumption for light duty vehicle, bus, and heavy duty truck platoons was investigated. For the light duty vehicle and bus platoons, the data available was direct force measurement. For the heavy duty truck platoons, the data available was fuel measurements. The fuel measurements were used to compute the drag forces using the VT-CPFM model. Subsequently, models were developed that capture the impact of the vehicle inter-platoon position and distance gap on the drag coefficient using a general power function.

The critical distance gap, the gap at which the drag coefficient is not affected by the vehicle ahead of it, is determined through optimization. It was demonstrated that the inclusion of this parameter in the optimization reduces the residual error (increases the fit accuracy) and yields a value that is consistent with empirical data. The model for the two-HDT platoon was validated against the drag coefficient from the numerical simulation results and fuel savings against in-field experimental measurements from the literature. For the drag coefficient, a high level of agreement was observed for the lead truck with some deviation for the trail truck. This disagreement is consistent with other literature concluding that the numerical simulation of the fluid flow is not suitable for modeling the drag interaction between vehicles. For the fuel savings, good agreement was observed with the field data in the literature while the trail truck model was consistent with the empirical data that exhibited a decreasing trend over the full range of the distance gaps. The developed drag models were used to quantify the average fuel savings for different types of platoons. The results show a potential decrease in the fuel consumption directly proportional with the inter-platoon distance gap/time gap. For safety considerations, and data and control latencies such as vehicle mechanical system response and latency in communication between vehicles, the bus and heavy duty truck platoons show more potential in the platoon energy savings at longer time gaps when compared to the light duty vehicle platoons. This highlights the need for platooning control to be directed towards buses and trucks.

\section*{Acknowledgments}
This effort was funded by the Office of Energy Efficiency and Renewable Energy (EERE), Vehicle Technologies Office, Energy Efficient Mobility Systems Program under award number DE-EE0008209.



\bibliographystyle{elsarticle-num}

\bibliography{References_Automotive}

\end{document}